\documentclass[onecolumn,10pt,aps,pre,notitlepage,superscriptaddress,longbibliography]{revtex4-1}
\usepackage{amssymb}
\usepackage{amsmath}
\usepackage{graphicx}% Include figure files
\usepackage{bm}% bold math
\usepackage{color}
\usepackage{ulem}
\nocite{apsrev41Control} 
\newcommand{\eps}{\varepsilon}
\begin{document}

%\title{From Benjamin-Feir instability to focusing dam breaks in water waves}
\title{From modulational instability to focusing dam breaks in water waves}
%\title{{\color{blue} Focusing dam breaks in modulationally unstable water waves}}

\author{F\'elicien Bonnefoy}
\affiliation{\'Ecole Centrale de Nantes, LHEEA, UMR 6598 CNRS, F-44 321 Nantes, France}
\author{Alexey Tikan}
\affiliation{Univ. Lille, CNRS, UMR 8523 - PhLAM -
 Physique des Lasers Atomes et Mol\'ecules, F-59 000 Lille, France}
\author{Fran\c{c}ois Copie}
\affiliation{Univ. Lille, CNRS, UMR 8523 - PhLAM -
  Physique des Lasers Atomes et Mol\'ecules, F-59 000 Lille, France}
\author{Pierre Suret}
\affiliation{Univ. Lille, CNRS, UMR 8523 - PhLAM -
  Physique des Lasers Atomes et Mol\'ecules, F-59 000 Lille, France}
\author{Guillaume Ducrozet}
\affiliation{\'Ecole Centrale de Nantes, LHEEA, UMR 6598 CNRS, F-44 321 Nantes, France}
\author{Gaurav Prabhudesai}
\affiliation{Laboratoire de Physique de l'Ecole normale sup\'erieure, ENS,
Universit\'e PSL, CNRS, Sorbonne Universit\'e, Universit\'e Paris-Diderot, Paris, France}
\author{Guillaume Michel}
\affiliation{Sorbonne Universit\'e, CNRS, UMR 7190, Institut Jean Le Rond d’Alembert, F-75 005 Paris, France}
\author{Annette Cazaubiel}
\affiliation{Universit\'e de Paris, Universit\'e Paris Diderot, MSC, UMR 7057 CNRS, F-75 013 Paris, France}
\author{Eric Falcon}
\affiliation{Universit\'e de Paris, Universit\'e Paris Diderot,  MSC, UMR 7057 CNRS, F-75 013 Paris, France}
\author{Gennady El}
\affiliation{Department of Mathematics, Physics and Electrical Engineering, Northumbria University, Newcastle upon Tyne, NE1 8ST, United Kingdom}
\author{St\'ephane Randoux}
\affiliation{Univ. Lille, CNRS, UMR 8523 - PhLAM -
  Physique des Lasers Atomes et Mol\'ecules, F-59 000 Lille, France}
 \email{stephane.randoux@univ-lille.fr}

\date{\today}% It is always \today, today,
             %  but any date may be explicitly specified

\begin{abstract}
   We report water wave experiments performed in a long tank where we
   consider the evolution of nonlinear deep-water surface gravity waves
   with the envelope in the form of a large-scale rectangular barrier.  Our
   experiments reveal that, for a  range of initial parameters, the
   nonlinear wave packet is not disintegrated by the Benjamin-Feir
   instability but exhibits a specific, strongly nonlinear modulation, which
   propagates from the edges of the wavepacket towards the center with finite speed.
   Using numerical tools of nonlinear spectral analysis of experimental data we
   identify the observed envelope wave structures with focusing
   dispersive dam break flows, a peculiar type of dispersive shock
   waves recently described in the framework of the semi-classical
   limit of the 1D focusing nonlinear Schr\"odinger equation
   (1D-NLSE). Our experimental results are shown to be in a  good
   quantitative agreement with the predictions of the semi-classical
   1D-NLSE theory. This is the first  observation of the  persisting
   dispersive shock wave  dynamics in  a modulationally unstable water
   wave system.
  \end{abstract}

%\pacs{Valid PACS appear here}% PACS, the Physics and Astronomy
                             % Classification Scheme.

\maketitle

\section{Introduction}
Following the pioneering works by Whitham and Lighthill \cite{Whitham:65,Lighthill:65},
Benjamin and Feir reported in 1967 the fundamental experimental and theoretical
investigations of the time evolution of nonlinear deep-water surface gravity waves
\cite{BF:67a,BF:67b}.  They demonstrated that a uniform continuous
wave train is unstable with respect to small long-wave perturbations of its envelope,
which may eventually  lead to its disintegration after some evolution
time~\cite{BF:67a,BF:67b,Lake:77,Longuet:80,Melville:82,Su:82,Su:82b}.  In 1968 Zakharov showed that, for
narrowband perturbations, the governing hydrodynamic equations can be
reduced to a single equation for the complex wave envelope: the
focusing one-dimensional nonlinear Schr\"{o}dinger equation (1D-NLSE)
\cite{Zakharov:68,Benney:67}. It was then understood that the instability, first
observed in water waves, represents a ubiquitous phenomenon in
focusing nonlinear media. Nowadays this phenomenon is called
modulational instability (MI) and it has been observed and studied in
many physical situations including plasma waves, matter waves, electromagnetic and optical waves
\cite{Zagryadskaya:68,Ostrovskii:72,Zakharov:09a,Strecker:02,Kip:00,Tai:86,Soljacic:00,Solli:12,Sun:12,Meir:04,Mosca:18}. 

According to the conventional picture, the early (linear) stage of MI is manifested in
the exponential growth of all the perturbations of a plane wave background that fall in
the region of the Fourier spectrum below a certain cut-off wavenumber.
  This simple classical (linear) picture provides a valid description
  of the process of the short-term destabilization of a plane wave of an infinite extent but it is
  inherently not adapted to the description of the nonlinear development of MI.
  Three distinct scenarios of nonlinear evolution of modulationally unstable wave systems
  described by the 1D-NLSE can be distinguished depending on the type of the initial condition considered.
  In the classical configuration {\it (i)} where the initial condition is a plane wave of infinite extent
  a particular scenario of the MI development strongly depends on the type of 
  perturbation of the plane wave background that is considered. Breather solutions of
  the focusing 1D-NLSE  are usually found to dominate the dynamics in this situation
  \cite{Erkintalo:11,Kibler:10,Frisquet:13,Akhmediev:11,Kimmoun:16,Mussot:18,Zakharov:13,Gelash:14,Kibler:15}
  although other nonlinear wave structures are also found depending on the localization and the solitonic content of the
  considered perturbation \cite{Biondini:16a,Biondini:16b,Biondini:17b,Kraych:19a,Conforti:18}.
  The destabilization of an infinite plane wave by a random perturbation leads to
  the emergence of a complex nonlinear wave structure associated  with the so-called
  integrable turbulence \cite{Zakharov:09} requiring  statistical approaches  to the
  description of the evolution of the nonlinear wave system  \cite{Randoux:14,Walczak:15,Agafontsev:15,Pelinovsky:06,Soto:16,Suret:16,Randoux:17,Tikan:18,Cazaubiel:18,Kraych:19b}. 

  There is another situation {\it (ii)} of physical relevance where the initial condition
  does not represent a plane wave of infinite extent but is a broad localized wavepacket
  with a smooth envelope. In this situation, the initial evolution is
  dominated by nonlinear effects, and the classical MI (understood as
  an exponential growth of small long-wave initial perturbations) plays a secondary role.
  The dynamical evolution of such wave fields in nonlinear focusing dispersive media
  gives rise  to generic dynamical features which are in sharp
  contrast  with the conventional MI scenarios \cite{Bertola:13}. As shown in the
  optical fiber experiments reported in ref. \cite{Tikan:17}, the
  nonlinear focusing of such wave packets (light pulses) results in a
  gradient catastrophe that is regularized by dispersive effects
  through the universal mechanism yielding the local Peregrine soliton
  structure \cite{Bertola:13}.  Note that these dynamical features
  have been observed in the nonlinear evolution of deep-water wave
  packets \cite{Shemer:02,Chabchoub:13} even though they had not been connected to
  the universal semi-classical mechanism of the generation of the
  Peregrine soliton discussed in ref. \cite{Bertola:13}. 

  In a third configuration {\it (iii)}
  which is the main focus of the present paper the initial field profile is characterized by sharp
  and significant amplitude changes. This configuration, which belongs to the class of the so-called Riemann
  problems \cite{Biondini:18}, can give rise to dispersive shock waves (DSWs), a phenomenon that has attracted
  considerable attention in recent years but considered predominantly for stable media \cite{El:16}.
  DSWs  in {\it shallow water waves} (often termed  {\it undular bores} ) are a classical subject of fluid
  dynamics \cite{whitham} with numerious contributions over the last 60 or so years \cite{Peregrine:66},
  starting from the pioneering paper by Benjamin and Lighthill
  \cite{Benjamin:54}. Note that some recent optical fiber
  experiments have demonstrated that the light may evolve as a fluid, mimicking
  the features of undular bores or dispersive dam break flows in shallow
  water \cite{Fatome:14,Xu:16,Xu:17}. 
  In contrast, there has been no experimental demonstration of
  the DSW dynamics on {\it deep water} so far.

In this paper, we present  water wave experiments in which we
demonstrate the persistent ``focusing DSW'' dynamics in the evolution
of water wave packets of large extent and constant amplitude.
Performing experiments in a long tank, we consider the evolution of
nonlinear deep-water surface gravity waves having their envelope in
the form of a large-scale rectangular barrier (a ``box'') of finite
height.  Our experimental observations reveal that, for a range of
input parameters,  the nonlinear wave train does not get disintegrated
by the spontaneous MI but instead,   exhibits
a regular DSW type behavior  that dominates the dynamics of the
nonlinear wave at intermediate times.  

More specifically, we observe that the initial sharp transitions
between the uniform plane wave and the zero background undergo a very
special nonlinear evolution leading to the emergence of two
counter-propagating focusing dispersive dam break flows  having the
characteristic DSW structure, in good agreement with the scenario
studied theoretically in ref. \cite{GEl:16,Jenkins:14} using
semi-classical analysis of the focusing 1D-NLSE. We show that there
exist ranges of parameters for which the dynamics observed in the
experiment are nearly integrable and quantitatively agree with the
theoretical predictions of \cite{GEl:16,Jenkins:14}.  We also show
that the observed behaviors exhibit significant degree of robustness
to perturbative higher-order effects. Our paper presents the
first observation of DSWs in deep water waves, supported by the
previously developed semi-classical theory \cite{GEl:16,Jenkins:14}.

The paper is organized as follows. In Sec. \ref{water_exp}, we describe our
experimental results obtained in a one-dimensional water tank. 
In Sec. \ref{SCth}, we introduce the semi-classical formalism
in which the observed dynamical behaviors can be interpreted.
In Sec. \ref{data_analysis}, we perform quantitative comparison between experimental
results and the semi-classical theory. In Sec. \ref{higher_order_effects}, we show that the observed
dynamics exhibits some degree of robustness to perturbative higher-order effects.
A brief summary of our work is presented in Sec. \ref{conclusion} together with
a short discussion about possible perspectives. 

%\newpage

\section{Water wave experiment}\label{water_exp}

\begin{figure}[h]  
  \includegraphics[width=1\textwidth]{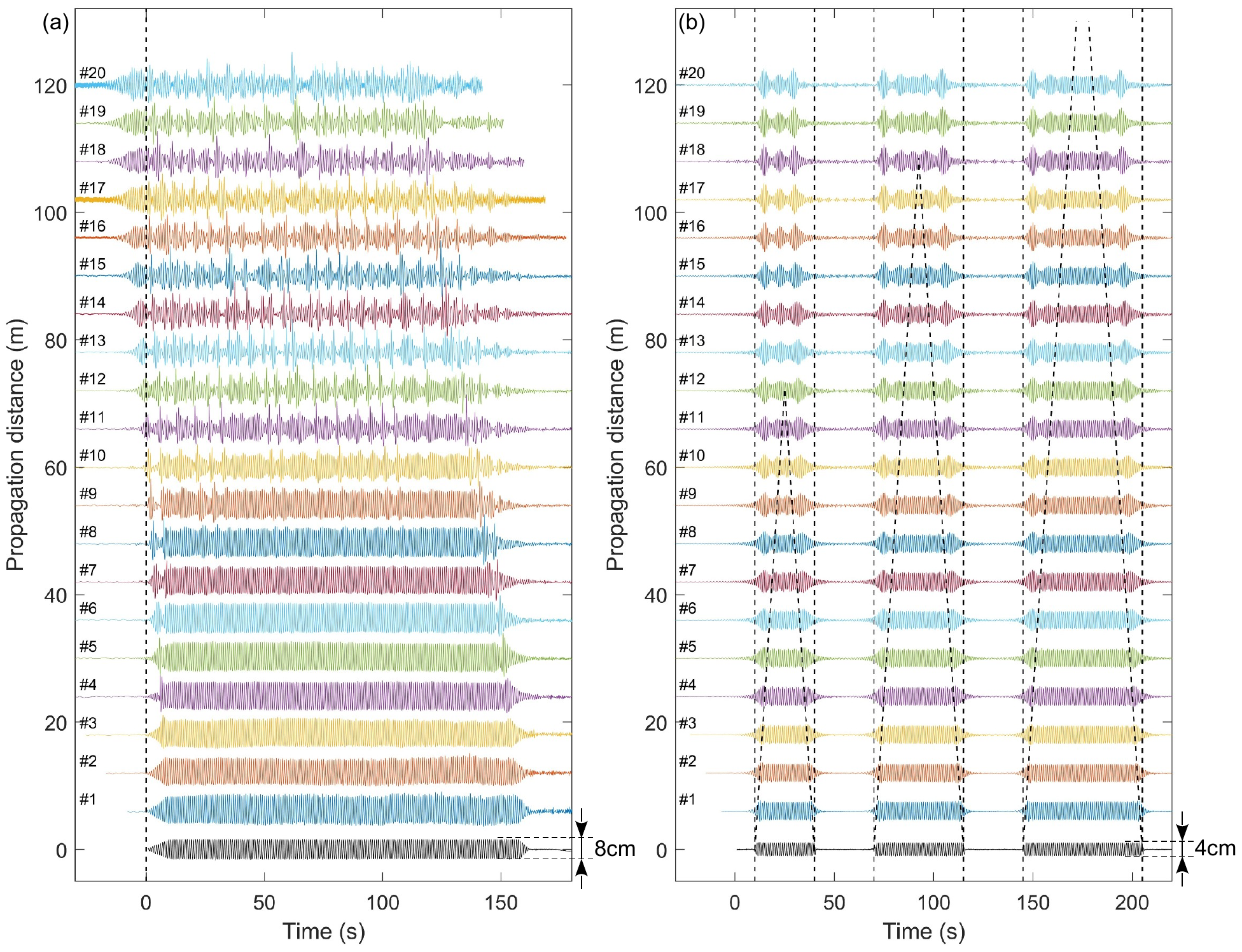}
  \caption{ Experimental results showing the nonlinear evolution of several rectangular wave trains along the
    1D water tank. The evolution is plotted in the frame of reference moving at the
    group velocity $\omega_0/(2k_0)$ of the wave packets. (a) A large scale wave packet of constant
    amplitude unstable to small perturbations of its envelope is disintegrated
    by the Benjamin Feir instability (The wave steepness is $k_0 a_0 \simeq 0.19$, the carrier period
    is $T_0=0.87$ s and the wave amplitude is $a_0=3.7$ cm (see text)). (b) Three ``boxes''
    of constant identical amplitudes, which are not disintegrated by the Benjamin-Feir instability, 
    undergo some strongly nonlinear modulation which propagates with finite speed from the 
      front and back edges of the
    wavepacket towards the center  under the form of counter-propagating dispersive
    dam break flows with DSW structure (The wave steepness is $0.082$, the carrier period
    is $T_0=0.99$ s and the wave amplitude is $a=2$ cm (see text)). The thick black dashed lines
    represent the theoretical breaking lines separating the genus $1$ region from the genus $0$ region, see
    calculation details in Sec. \ref{semi_class_results} and Sec. \ref{cnoidal}.
  }
  \label{fig1}
\end{figure}

The experiment was performed in a wave flume at the Hydrodynamics,
Energetics and Atmospheric Environment Lab (LHEEA) in Ecole Centrale de Nantes
(France). The flume which is $148$ m long, $5$ m wide and $3$ m deep is equipped
with a parabolic shaped absorbing beach that is approximately $8$ m long. With the
addition of pool lanes arranged in a $W$ pattern in front of the
beach the measured amplitude reflection coefficient is as low as $1\%$.
Unidirectional waves are generated with a computer assisted flap-type wavemaker. 
The setup comprises $20$ equally spaced resistive wave gauges that
are installed along the basin at distances $z_j=6+(j-1)6$ m, $j=1,2,...20$
from the wavemaker located at $z=0$ m. This provides an effective measuring range
of $114$ m.

In the first experimental run presented in Fig. 1(a), the wavemaker
produced one single large-scale wavepacket having a near rectangular envelope.
The duration $\Delta T_0$ of the wave packet is $\sim 160$ s.  
The water wave has a carrier period $T_0=2\pi/ \omega_0$ of $0.87$ s. The angular frequency
$\omega_0$ of the wave and the wave vector $k_0$ are linked according to the deep
water dispersion relation $\omega_0^2=k_0 g$  ($k_0=5.31$ m$^{-1}$, $\lambda_0=2\pi/k_0\simeq1.17$ m) 
with $g$ the gravity acceleration. The amplitude of the generated
envelope is $a_0=3.7$ cm implying that the wave steepness $k_0 a_0$ is $\simeq 0.19$
in this experiment. 

In the second experimental run presented in Fig. 1(b), the computer controlled wavemaker
produced a sequence of three large-scale wave packets having 
rectangular envelopes. The three rectangular wave trains are individually
generated over a global time interval of $\sim 220$ s where they have increasing
durations of $\Delta T_1 = 30$ s, $\Delta T_2 = 45$ s, $\Delta T_3 = 60$ s. The
period of the carrier wave has been changed to $T_0=0.99$ s ($k_0=4.10$ m$^{-1}$, $\lambda_0=2\pi/k_0\simeq1.51$ m)
and the amplitude of the generated envelope has been reduced to $a=2$ cm
which implies that the wave steepness $k_0 a = 0.082$ is $2.3$ times smaller than in Fig. 1(a). 

 Fig. 1(a) shows that disintegration of the large rectangular
wavepacket occurs as a result of modulational instability, in good
agreement with experimental results previously reported in ref. \cite{BF:67b,Lake:77,Su:82,Su:82b,Chabchoub:17}. 
In particular, the initial destabilization of the plane wave
background ($z<40$ m) is associated with the exponential amplification of
small random perturbations having frequency components falling in spectral
regions of MI gain. We have checked that the amplification  rate
measured in our experiment is in agreement with the theory developed
by Benjamin and Feir \cite{BF:67a,BF:67b,Cazaubiel:18}. At long propagation distances ($z>50$ m),
the spectral (Fourier) analysis of the experimental wave train reveals
a significant frequency down-shifting ($\sim 0.15$ Hz at $z=120$ m) correlated with a
spectral broadening (see Fourier spectra plotted in the Supplemental
Material \cite{note_supl}). These spectral features
already observed in ref. \cite{Su:82} demonstrate that the dynamical evolution
reported in Fig. 1(a) is influenced by high-order nonlinear
effects. Another clear signature of the presence of higher-order nonlinear effects
lies in the fact that the front edge and the back edge of the wavetrain in Fig. 1(a)
are propagating faster than the reference frame moving at the group velocity $\omega_0/(2k_0)$,
a feature which is not observed in Fig. 1(b).
Such a space-time evolution and the observed frequency down-shifting cannot be
described by the 1D-NLSE but rather by other models like the
unidirectional Zakharov equation or the Dysthe equation. The
occurrence of higher-order nonlinear effects breaks the integrability
of the wave system (see also Fig. 7(a)(b)(c)(d)) and it prevents the
observation of any recurrence phenomenon, which represents an
intrinsic feature of integrable wave systems.

%Fig. 1(a) shows that disintegration of the large rectangular wavepacket
%occurs as a result of Benjamin-Feir instability, in good qualitative agreement with
%experimental results previously reported in ref. \cite{BF:67b,Lake:77,Su:82,Su:82b,Chabchoub:17}. 
%Note that some dispersive breaking starts to occur near the edges of
%the box at a propagation distance of $\sim 42$ m, simultaneously
%with the growth of the Benjamin-Feir instability which is observable
%around the center of the large box. It is however clear
%that the Benjamin-Feir instability dominates the wave dynamics for the 
%value of the steepness ($k_0a_0 \simeq 0.19$) associated with the experiment
%shown in Fig. 1(a). 

Fig. 1(b) reveals that features of a qualitatively different nature
occur at smaller steepness and for the rectangular wavepackets
having shorter durations. 
Each of the three generated rectangular wave packets
qualitatively experiences some dispersive breaking following a scenario 
where two nonlinear wave trains generated from the edges of the rectangular
envelopes counterpropagate towards the center of the box envelopes, see also the
Supplemental Material Video S1 \cite{note_supl}.
These two counterpropagating nonlinear wave trains were identified as two
dispersive dam break flows in ref. \cite{GEl:16}. 
Remarkably, the relatively moderate steepness ($k_0a \simeq 0.08$) characterizing
water waves used in our experiment permits the clear observation
of this phenomenon without the Benjamin-Feir
instability significantly perturbes the wave dynamics. 

Let us mention that some features qualitatively similar to those shown in Fig. 1(b) have been reported
in ref. \cite{Hwung:07,Tulin:99} for water waves. In these experiments
it was understood that the evolution of the unsteady wavefront
was determined by combined influence of nonlinearity and dispersion
but the degree of the analysis that was made did not exceed a very qualitative
level.

In 2015, Shemer and Ee have reported some experiments
showing the evolution of a truncated water wave train having
a rectangular shape at initial time \cite{Shemer:15}. The physical values of
parameters used in their experiments ($T_0=0.8$ s, $a=2.6$ cm) are close to those
used in the experiments that we report in Fig. 1(b). However the truncated wave
train considered by Shemer and Ee was weakly modulated
in such a way that a Peregrine breather was generated after some
propagation distance inside the water tank \cite{Shemer:15}.
Even though the growth of large oscillations at the edges of the
truncated wave train was reported, most of the study presented
in ref. \cite{Shemer:15} was focused on the build-up of the
Peregrine breather generated in the central part of the rectangular
wave train. 

Interestingly our water wave experiments can be connected to the subject
of diffractive focusing of waves in time and in space. In ref. \cite{Weisman:17},
Weisman {\it et al} have reported an experiment where the envelope of a surface
gravity water wave is modulated in time by a rectangular function. Near-field
(Fresnel) diffraction patterns very similar to those observed in optics for
light beams diffracted by a slit have been observed in the water wave context. 
Contrary to our experiment, the experiments reported in ref. \cite{Weisman:17} 
are placed in a purely linear regime where (linear) diffraction of waves is
observed. As it is clearly shown in Sec. \ref{data_analysis},
our experiments involve nonlinear wavefields
that have some solitonic content. From an optical perspective, they are
conceptually related to the subject of nonlinear diffraction of a field of constant
amplitude by a slit in a focusing medium \cite{Akhmanov:68}, a research topic introduced 
at theoretical level by Manakov \cite{Manakov:74a,Manakov:74b}.

Nonlinear diffraction of light beams in focusing media has been
considered in a few optical experiments.  The experiment reported in
ref. \cite{Wan:10} has investigated diffraction from an edge in a
self-focusing nonlinear photorefractive medium using a spatially
incoherent light beam. In the very recent experimental
work \cite{Marcucci:19} the evolution of a 1D optical beam having a square profile was
observed in a focusing photorefractive medium. While some of the robust
qualitative  features of the DSW dynamics  predicted by the
semi-classical 1D-NLSE theory \cite{El:16} have been observed
and interpreted in the context of the ``topological control of extreme
waves'' \cite{Marcucci:19}, the quantitative comparison with the theory was
limited because of the significant competition between the DSW dynamics
and noise amplification in the modulationally unstable photorefractive
medium.

Recent optical fiber experiments reported in ref. \cite{Audo:18} have also evidenced
a spatio-temporal evolution very similar to the one that we observe with
the rectangular wave train of the smallest width ($\Delta T_1 = 30$ s),
compare Fig.1 (b) with Fig. 3(a) of ref. \cite{Audo:18}. However
the work reported in ref. \cite{Audo:18} was concentrated on the
emergence of Peregrine-like events and did not allow for a
meaningful quantitative, or even qualitative, identification of the
observed wave patterns with DSWs due to very few oscillations observed.

\section{Dispersive focusing dam break flows: semi-classical theory}\label{SCth}

The experimental results  shown in Fig.~\ref{fig1}(b)
clearly indicate  that the envelope of the wavepacket develops
oscillations with  the typical period  significantly smaller than
the temporal extent of the wavepacket.  This separation of scales
suggests the  usefulness of an asymptotic WKB-type approach to the
theoretical understanding of the arising dynamics. In this section,
we show that the mathematical framework of dispersive
hydrodynamics \cite{Biondini:16}, a semi-classical theory of
nonlinear dispersive waves,  provides some insightful interpretation of the experimentally
observed multi-scale coherent structures.

\subsection{The semi-classical framework}\label{semi_class}

\begin{figure}[h]
  \includegraphics[width=1\linewidth]{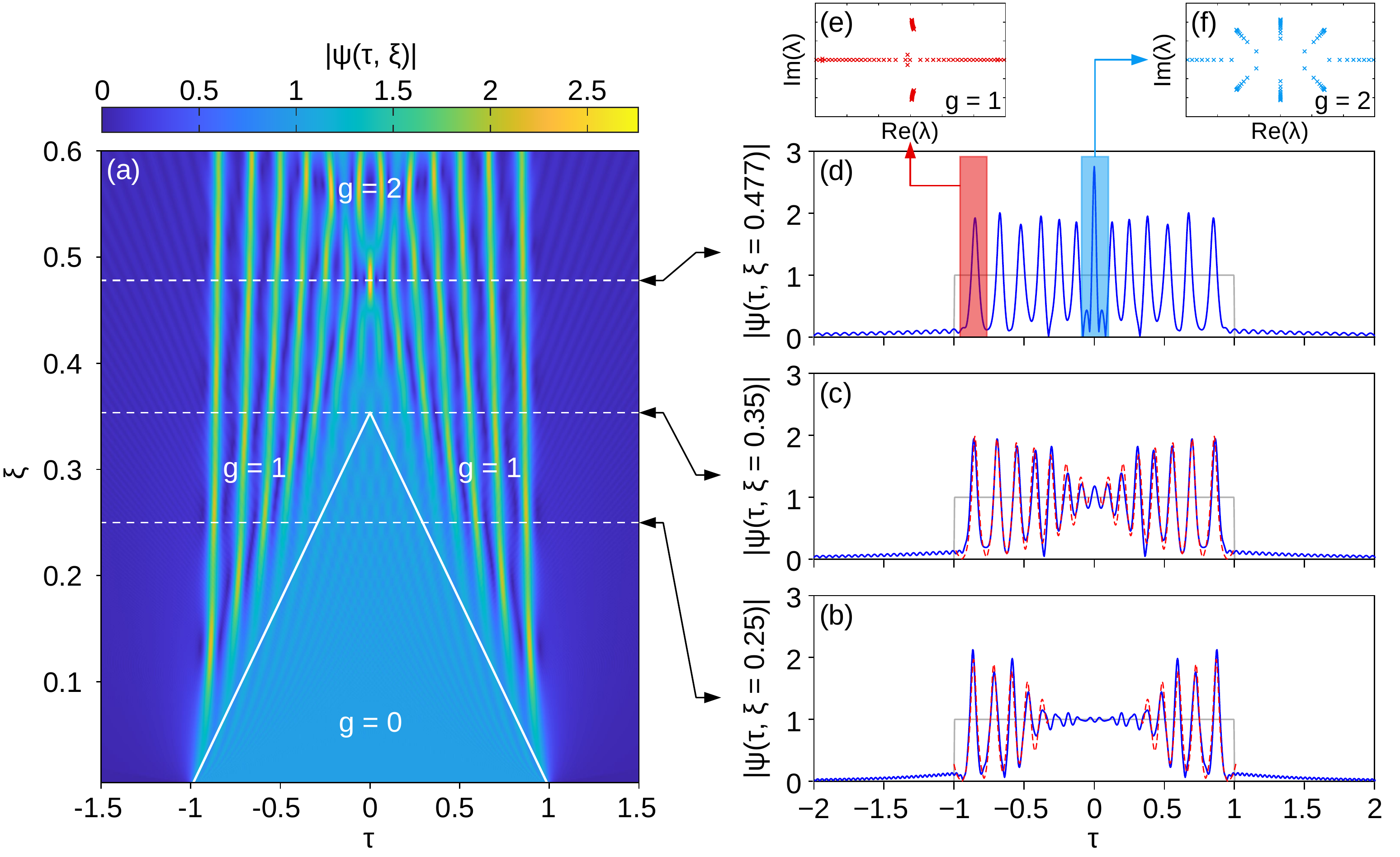}
  \caption{Numerical simulation of Eq. (\ref{nlse}) showing (a) the space
    time evolution of the wave field having a profile specified by Eq. (\ref{ic0}) at $\xi=0$
    ($q=1$, $T=1$, $\varepsilon=0.04$). The space-time evolution
    is separated into three regions of increasing genus $g$ (see text). The genus $g=0$ region
    corresponds locally to the plane wave solution. The genus $g=1$ region
    is associated to DSWs that are generated from the edges of the box.
    The genus $g=2$ region emerges from the collision of the two focusing dam break flows
    , see also the Supplemental Material Video S2 \cite{note_supl}.
    Curves plotted in blue lines
    in (b), (c), (d) represent the wave amplitude profiles at $\xi=0.25$, $\xi=0.35$, $\xi=0.477$,
    respectively. Red dashed lines in (b) and (c) represent the amplitudes of the modulated cnoidal
    waves that are determined from Eqs. (\ref{elliptic}), (\ref{mod1}), (\ref{mod2}).
    (e), (f) Spectral (IST) portraits of isolated structures made at $\xi=0.477$.
    The spectral portrait in (e) is mostly composed of two complex conjugate bands
    demonstrating that the analyzed structure has a genus $g=1$ (soliton-like structure,
    see text).
    The spectral portrait in (f) is  composed of three  bands
    demonstrating that the genus of the analyzed structure is $g=2$ (breather-like structure). } \label{fig2}
\end{figure}

%The relations between the parameters in Eqs.~\eqref{nlse} and \eqref{nlse_phys} are given by Eq.~\eqref{math_phys_var}.

The experimental results reported in Fig. 1(b) can be interpreted within
the framework of the focusing 1D-NLSE \eqref{nlse}, written in a
dimensional form as a spatial evolution equation  
\begin{equation}\label{nlse_phys}
i \frac{\partial A}{\partial z} + \frac{k_0}{\omega_0^2}
\frac{\partial^2 A}{\partial t^2} + \alpha k_0^3 |A|^2 A=0 ,  
\end{equation}
where $A(z,t)$ represents the complex envelope of the water wave
that changes in space $z$ and in time $t$ \cite{osborne2010nonlinear}.
$\alpha=0.91$ is a corrective term to the cubic nonlinear term.
It has been introduced in Eq. (\ref{nlse_phys})
in order to take into account finite depth effects. In our experiment
where the water depth $h$ is $3$ m, the numerical value of  $k_0 h$ is
$ \sim 12.3$. This is large enough to consider that the condition of
propagation in deep water regime is well verified but not large enough not to
include some small corrective term in the nonlinear coefficient.
A comprehensive discussion about the influence of finite depth effects
on the values of linear and nonlinear coefficients is given
in Appendix \ref{appendixa}.

In the experimental evolution reported in Fig. 1(b), the dynamics of the nonlinear wave
is ruled by the interplay of two characteristic length scales associated
with the temporal duration $\Delta T_j$ $(j=1,2,3)$ of the rectangular envelopes,
namely, the nonlinear
length $L_{NL}=1/(\alpha k_0^3 a^2)$ and the linear dispersion length
$L_D=(\omega_0 \Delta T_j)^2/(2 \, k_0)=g \Delta T_j^2/2 $. Normalizing the propagation
distance $z$ along the flume as $\xi=z/\sqrt{L_{NL}L_D}$, the physical
time as $\tau=t/ \Delta T_j$, the complex field envelope as $\psi=A/a$,
Eq. (\ref{nlse_phys}) takes the following dimensionless ``semi-classical'' form

\begin{equation}\label{nlse}
i \epsilon \frac{\partial \psi}{\partial \xi} + \frac{\epsilon^2}{2}
\frac{\partial^2 \psi}{\partial \tau^2} +|\psi|^2 \psi=0
\end{equation}
where $\epsilon=\sqrt{L_{NL}/L_D} \ll 1$ is a small dispersion parameter. 

The numerical values of the physical and dimensionless parameters describing
our experiment are reported in Table I for the three rectangular wave trains with
temporal widths $\Delta T_j$ ($j=1,2,3$). It can be easily seen that our experiments
are always placed in a regime where $L_{NL} \ll L_D$ which implies that
the experimental values of the $\epsilon$ parameter are much smaller than $1$.
Therefore our experimental observations can be interpreted within
the mathematical framework of dispersive hydrodynamics \cite{Biondini:16},
a semi-classical theory of nonlinear dispersive waves suitable for 
such multi-scale coherent structures.

\begin{table}[h]\label{table1}
  \centering
  \begin{tabular}{p{13mm}p{13mm}p{13mm}p{13mm}p{13mm}p{13mm}}
\hline%
$\Delta T_j$ & $L_{NL}$ & $L_D$ & $\epsilon $ & $ z_j^* $& $ N $ \\
( s ) & ( m ) & ( m ) & &(m) &  \\\hline %
30 & 39.86 & 4414 & 0.095 & 74 & 3 \\\hline %
45 & 39.86 & 9932 & 0.063 & 111 & 5 \\\hline %
60 & 39.86 & 17658 & 0.047 & 148 & 7 \\\hline %
\end{tabular}
  \caption{Parameters corresponding to the three rectangular
    wave envelopes considered in our experiment ($k_0=4.1$ m$^{-1}$, $\omega_0=6.34$ s$^{-1}$,
    $a=2.\, 10^{-2}$ m, $\alpha=0.91$). $\epsilon$ is the small
    dispersion parameter in Eq. (\ref{nlse}). $z_j^*$ represents the physical position
    at which the dispersive dam break flows collide. $N$ represents the number of
  solitons embedded within the rectangular wavepackets, see Sec. \ref{ist}.}
\end{table}

Eq. \eqref{nlse}  is considered with  decaying data in the form of a
rectangular barrier  of finite height $q>0$  and the width $2T$: 
\begin{equation}\label{ic0}
\psi(\tau,\xi=0) =\left\{
\begin{array}{ll}
q  &\quad \hbox{for} \quad |\tau| <T,\\ 0& \quad \hbox{for} \quad |\tau|>T
\, .
\end{array}
\right.
\end{equation}
We shall call the initial value problem \eqref{nlse}, \eqref{ic0} the
1D-NLSE box problem.

Fig. 2(a) shows the numerical simulation of the
1D-NLSE box problem for $\epsilon=0.04$, see
also the Supplemental Material Video S2 \cite{note_supl}. Fig. 2(a) provides evidence of
a space-time evolution qualitatively similar to the one observed in our
water wave experiment, see Fig. 1(b). In particular the two nonlinear wave trains
that are generated at the edges of the box counterpropagate towards
the center of the box where they collide and produce a peak of large
amplitude, see Fig. 2(d). Let us emphasize that
  simulations reported in Fig. 2 are done for an unperturbed box (see Eq. (\ref{ic0})).
  If the box is perturbed by a small modulation or a small noise having frequency components
  falling within the MI gain curve, the perturbation will be amplified and an interaction
  between the DSWs generated from the edges of the box and the coherent structures seeded
  by the initial perturbation will be
  observed. The study of the associated dynamics represents an interesting perspective
  of our work.

Note that the typical scale of the coherent structures found in the dam break flows
is around $\sim \epsilon \ll T$, see Fig. 2(b),(c),(d). Note also that
  the dynamics reported in Fig. 2(a) is not influenced by the exact shape of the field
  near the edges of the box. Space-time evolutions similar to the one reported in Fig. 2(a)
  are observed as long as the typical space scale of the transition between the zero and the constant
  backgrounds is of the order of $\epsilon$.

\subsection{Main results from the semi-classical theory}\label{semi_class_results}

In this section, we summarize some important theoretical results about
the 1D-NLSE box problem. The quantitative comparison between these theoretical
results and the experimental results will be presented in Sec. \ref{data_analysis}.

First, it is instructive to use the Madelung transform
$\psi=\sqrt{\rho(\tau,\xi)} \exp(i \epsilon^{-1}\int^\tau u(\tau',\xi) d\tau')$ to represent the
1D-NLSE in the dispersive hydrodynamic form 
\begin{equation}\label{dh}
\begin{split}
    \rho_\xi+(\rho u)_\tau=0,\\ u_\xi+uu_\tau - \rho_\tau -
    \eps^2\left(\frac{\rho_\tau^2}{8\rho^2}
    -\frac{\rho_{\tau\tau}}{4\rho}\right)_\tau=0 \, ,
   \end{split}
\end{equation}
where $\rho$ and $u$ are analogues of the fluid depth and  velocity
respectively.  Within the hydrodynamic interpretation [Eq. \eqref{dh}], the
box initial data [Eq. \eqref{ic0}] can be viewed as a combination of two
hydrodynamic dam breaks (i.e. step transitions from finite depth
$\rho=q^2$ to ``dry bottom'' $\rho=0$) of opposite polarities, placed
at the distance $2T$ from each other.  It is important to stress that,
the 1D-NLSE ``fluid'' here has nothing to do with the underlying water
wave context of the original problem; moreover, due to the focusing
nature of the 1D-NLSE \eqref{nlse}, the classical ``pressure'' term in
the hydrodynamic representation \eqref{dh} is negative. 

The dispersive hydrodynamic representation \eqref{dh} provides an
important insight into the  1D-NLSE evolution of different types of initial
data. Linearising system \eqref{dh} about a constant equilibrium flow
$\rho=\rho_0$, $u=0$ (a plane wave of the 1D-NLSE) one obtains the
usual 1D-NLSE dispersion relation $\omega=\pm k\sqrt{(\eps
  k)^2-4\rho_0}$ implying modulational  instability of plane waves for
long enough waves with $\eps k< 2\sqrt{\rho_0}$. This is the classical
Benjamin-Feir instability, which is manifested as a
dispersion-dominated, linear wave phenomenon {\it within 1D-NLSE}. The
initial exponential growth of harmonic, long-wave perturbations is
mediated by nonlinearity leading to the formation of Akhmediev
breathers or more complicated breather structures associated
with integrable turbulence \cite{Dudley:14, Randoux:16a}.

The focusing 1D-NLSE dam break problem \eqref{nlse}, \eqref{ic0} is
rather special in the sense that it triggers both nonlinearity and
dispersion in \eqref{dh} from the early time of the evolution. As a
result, it leads to the formation of a coherent, unsteady nonlinear
wave structure that is very different from those arising in the
development of the BF instability or in the evolution of broad smooth
humps. This  structure can be viewed as a focusing counterpart of the
well-known dispersive-hydrodynamic phenomenon, called a dispersive
shock wave (DSW) \cite{El:16}, which represents an expanding,
nonlinear wave train connecting two disparate  constant fluid
states. DSW is described by a slowly modulated, locally periodic wave
solution of a dispersive equation (1D-NLSE in our case) gradually
transforming from a soliton at one edge to a vanishing amplitude,
harmonic wave at the opposite edge. The special modulation providing
such a transition has been found in \cite{GEl:93} as a  self-similar
solution of the  Whitham modulation equations \cite{whitham} associated with
the 1D-NLSE.  Typically, DSWs are the features of stable media,
described by such equations as the KdV or defocusing NLS equations
(see \cite{El:16} and references therein) but for a special Riemann
data (dam break) the DSWs can be generated in unstable (focusing)
media (\cite{Kamchatnov:97}, \cite{GEl:16}, \cite{Biondini:17}). The
persistence of  DSW dynamics in focusing dam break problem is due to a
special ``hyperbolic'' modulation as explained below.

The  periodic solutions of the focusing 1D-NLSE are known to be
modulationally unstable with respect to small initial perturbations
(see e.g. \cite{Agafontsev_2016}), but this modulational instability is more subtle than the BF
instability of a plane wave. It turns out that the instability of
nonlinear periodic solution can be ``inhibited''  by a special
modulation yielding a ``hyperbolic''  wave behavior characterized by
finite speeds of propagation.  This modulation is described by a
similarity solution of the Whitham modulation equations associated
with the 1D-NLSE \cite{pavlov_nonlinear_1987}, \cite{forest_geometry_1986} and it is exactly
the modulation that is realised in the dispersive regularization of
the dam break flow in the focusing 1D-NLSE and enables the persistent
DSW structure that can be observed in an experiment.
  
The  box problem \eqref{nlse}, \eqref{ic0}  has been studied
analytically in \cite{GEl:16, Jenkins:14} using a combination of the Whitham
modulation theory and an IST-based Riemann-Hilbert problem approach \cite{tovbis_semiclassical_2016}.
The theoretical developments of \cite{GEl:16} important for  the
interpretation of our experimental results  in water waves can be
conveniently explained by considering Fig.~\ref{fig2}  where the numerical
simulation of the focusing dam break problem for the 1D-NLSE equation
is presented  along with the results of the so-called ``local IST''
analysis \cite{Randoux:16a}  of the emerging wave structures (Fig. 2(e),(f)).
The plots in Fig. 2(e),(f) show the qualitative
changes  of the nonlinear (IST) spectra occurring in the course of the
wave propagation.  These spectra and the associated nonlinear waves
are characterised by a fundamental integer index $g$ called genus
which enables classification of the emerging wave structures in terms
of the number $N=g+1$ of ``nonlinear Fourier  modes'' involved. The genus itself characterises topology of the hyperelliptic
Riemann surfaces associated with the special class of the 1D-NLSE
solutions, called finite-gap potentials (see e.g. \cite{belokolos_algebro-geometric_1994, osborne2010nonlinear}). As shown in
\cite{GEl:16} the solutions of the semi-classical 1D-NLSE box problem
can be asymptotically described by slowly modulated finite-gap
1D-NLSE solutions with the genus changing across certain lines in $\tau$-$\xi$
plane called {\it breaking curves}.   In particular,  the wave
structures regularising the initial dam breaks at $\tau=\pm T$ in the box
problem have genus $g=1$ while the genus $2$ structures emerge as a
result of the interaction of two counter-propagating  dispersive dam
break flows having the signature structure of dispersive shock waves
(DSWs) \cite{El:16}.  

The  asymptotic solution of  the box  problem for the small dispersion
1D-NLSE \eqref{nlse} has different form in different regions of $\tau$-$\xi$ plane
(see Fig.~\ref{fig2}). For $\xi<\xi^*$, where $\xi^* = \frac{T}{2
  \sqrt{2}q}$ the solution represents two
counter-propagating focusing DSWs---seen as the genus one regions in Fig.~\ref{fig1}---connecting two disparate genus
zero states: the ``dry bottom'' state $\psi=0$ at $|\tau|>T$ and the
constant state $\psi=q$ for  $(2 \sqrt{2} q \xi -T) <\tau< ( -2\sqrt{2} q
\xi+T) $. The local structure of both DSWs is described by the elliptic
(``cnoidal'') solution of the 1D-NLSE
\begin{equation}\label{elliptic}
\rho=(q+b)^2 - 4q b \, \hbox{sn}^2 \left( 2\sqrt{q b/m} \, (\tau - a \xi -
\tau_0)\varepsilon^{-1}; m \right),\end{equation} where
$\hbox{sn}(\cdot)$ is a Jacobi elliptic function with the modulus $ m
\in [0,1]$ given by
$$ m=\frac{4q b}{a^2+(q+ b)^2}\, .
$$ The modulation parameters $a(\tau, \xi)$, $b(\tau,\xi)$ are  found from
equations
\begin{equation}\label{mod1}
\begin{split}
a= \pm \frac{2q}{m \mu(m)}\sqrt{(1-m)[\mu^2(m)+m -1]} , \\ b =
\frac{q}{m \mu(m)}[(2-m)\mu(m) -2(1-m)] \, ,  
\end{split}
\end{equation}
\begin{equation}\label{mod2}
\begin{split}
\frac{\tau \mp T}{\xi}= \pm \frac{2q}{m \mu(m)}\sqrt{(1-m)(\mu^2(m)+m-1)}
\\  \times \left ( 1+ \frac{(2-m)\mu(m) -2(1-m)}{\mu^2(m) +m -
  1}\right) \, ,
\end{split}
\end{equation}
where   $\mu(m)=E(m)/K(m)$. $K(m)$ and  $E(m)$ are the complete
elliptic integrals of the first and second kind respectively.  The
signs $\pm$ in \eqref{mod1}, \eqref{mod2} correspond to right- and
left- propagating waves. The initial position $\tau_0$ in \eqref{elliptic} is
given by $\tau_0=\pm T$. In practice, $\tau_0$
depends on the way the sharp edges of the ``box'' are smoothed in the experimental
signal or in numerical simulations so for a practical comparison  with the theory,
one chooses $\tau_0$ by fitting to the experimental/numerical data.

Solution \eqref{elliptic}, \eqref{mod1}, \eqref{mod2} describes two
symmetric  oscillatory structures exhibiting the fundamental 1D-NLSE
solitons ($m=1$) with the amplitude $|\psi_m|=2q$, located at
$\tau=\pm T$. The structures degenerate, via the modulated elliptic
regime, into the vanishing amplitude linear wave ($m=0$) at the
internal moving  edges propagating towards the box centre with
constant velocities $\pm 2 \sqrt{2} q$. The solutions computed from
Eq. \eqref{elliptic}, \eqref{mod1}, \eqref{mod2} are plotted with
red dashed lines in Fig. 2(b),(c). Very good quantitative agreement
is found between these theoretical solutions and numerical simulations
of the 1D-NLSE (blue lines in Fig. 2(b),(c)). Let us
  recall, for the sake of clarity, that the theoretical and numerical solutions
  discussed here represent the envelopes (modulations) of wavepackets plotted
  in Fig. 1(b).

The equation of the first
breaking curve $\Gamma_1$ separating the genus $g=0$ region
from the genus $g=1$ region in the diagram in Fig.~\ref{fig2}(a)  is 
\begin{equation}\label{Xi1}
\Gamma_1: \qquad \xi= \frac{T-|\tau|}{2 \sqrt{2}q}  \, 
\end{equation}
Equation \eqref{Xi1} yields the DSW collision time $\xi^*= \frac{T}{2
  \sqrt{2}q}$ corresponding to Fig.~\ref{fig2}(c). For $\xi>\xi^*$ the region with $g=2$ is formed,
confined to another breaking curve (not shown in Fig.~\ref{fig2}(a)). One
of the prominent features of the genus 2 region is the occurrence of a
large-amplitude breather at the center with the characteristics close
to those of the Peregrine soliton  (see 
 Fig.~\ref{fig2}(d)) as predicted in \cite{GEl:16} and
experimentally observed in fiber optics in \cite{Audo:18}.

We now demonstrate that deep water waves, while providing the
classical example of the Benjamin-Feir instability, present also a
medium  supporting the ``hyperbolic'' dispersive dam break (DSW)
scenario of the wavepacket evolution for a range of input
parameters. This is done in Sec. \ref{data_analysis} by a quantitative comparison of the water
wave experiment with the modulated 1D-NLSE solution \cite{GEl:16} and
the ``local IST'' analysis  of the experimentally observed wave
patterns \cite{Randoux:16a, Randoux:18}, confirming  the spectral
topological index (genus) of  generated waves.

\section{Data analysis and comparison with the theory}\label{data_analysis}

\subsection{Numerical simulations of the 1D-NLSE, breaking lines, collision points and modulated cnoidal waves}\label{cnoidal}

In this section, we focus on the quantitative comparison between experimental
results and the semi-classical theory. 
First we have performed numerical simulations of Eq. (\ref{nlse_phys}) by
taking as initial condition the complex envelope $A(z_1,t)$ of the
signal measured by the gauge closest to the wavemaker ($z_1=6$ m).
The complex envelope has been computed from the experimentally-recorded
signals by using standard techniques based on the Hilbert
transform, as discussed e. g. in ref. \cite{osborne2010nonlinear}. 
Fig. 3 shows the modulus $|A(z_{20},t)|$ of the complex
envelope that is computed at $z_{20}=120$ m, the position where is located the
gauge furthest from the wavemaker. The agreement between the experimental
results and the numerical simulations is quantitatively  good for each
of the three generated rectangular wave trains.

\begin{figure}[h]
  \includegraphics[width=0.8\linewidth]{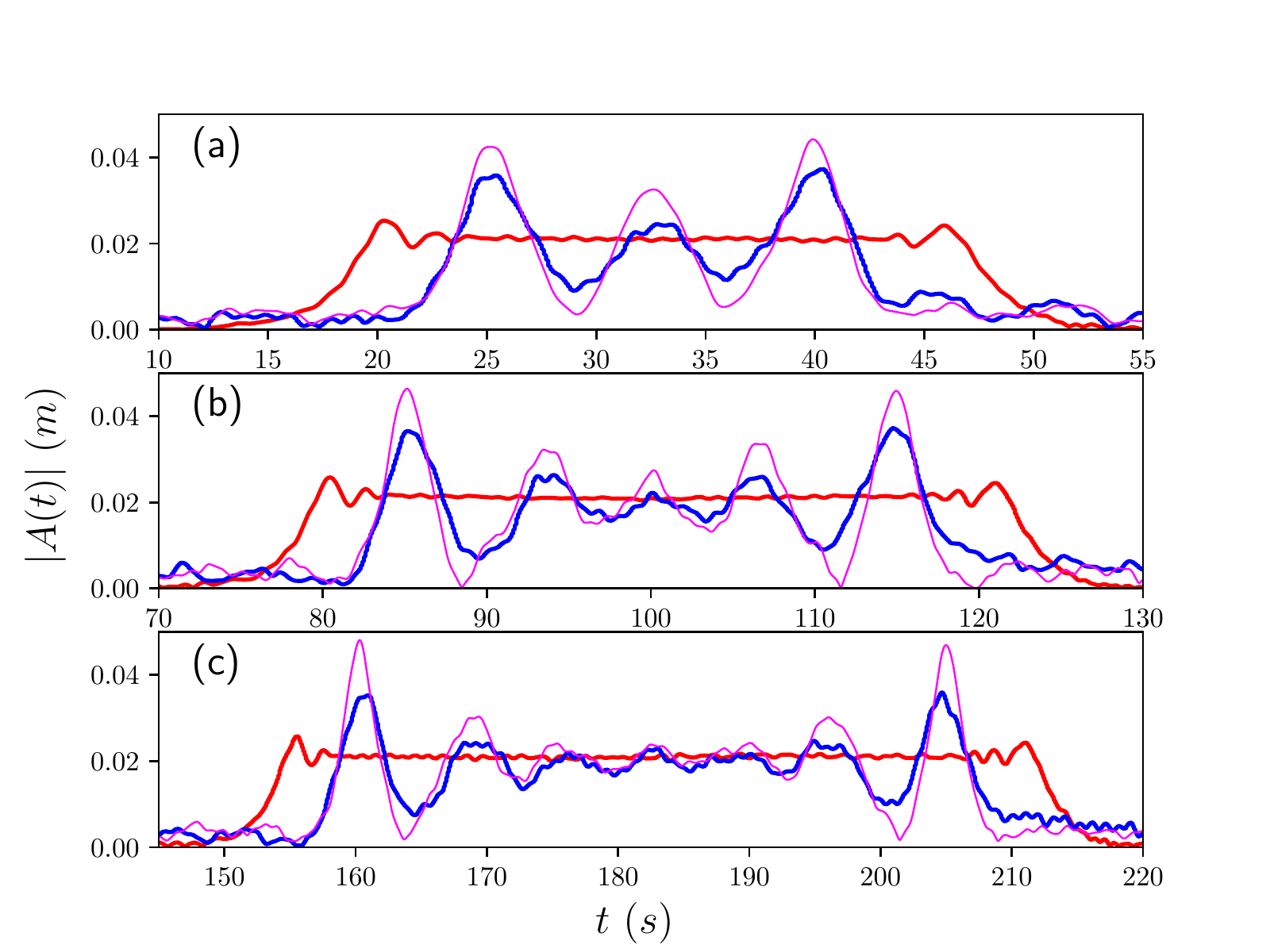}
  \caption{Modulus of the water wave envelopes
    with durations (a) $\Delta T_1=30$ s,
    (b) $\Delta T_2=45$ s, (c) $\Delta T_3=60$ s. The red (resp. blue) lines represent the
    experimental envelopes of the signal recorded at $z_1=6$ m (resp. $z_{20}=120$ m),
    close to (resp. far from) the wavemaker. 
    The magenta lines represent the envelopes computed from the numerical simulation
    of Eq. (\ref{nlse_phys}) ($k_0=4.1$ m$^{-1}$, $\omega_0=6.34$ s$^{-1}$, $\alpha=0.91$)
    by taking as initial condition the complex envelope measured by the
    gauge closest to the wavemaker (red lines, $z_1=6$ m).}\label{fig3}
\end{figure}

As a first valuable test of the theory introduced in Sec. \ref{SCth}, 
we plot the linear breaking curves separating the genus $0$ (plane wave)
regions from the genus $1$ (DSW) regions. Rephrasing Eq. (\ref{Xi1})
in physical units, we easily find that the slopes $s_\pm$ of the breaking lines
in the $z$-$t$ plane read $s_\pm = \pm \omega_0/(4\, a \, \sqrt{\alpha} \, k_0^2)$
and that the collision between the two counterpropagating dam break flows
occurs at the position $z_j^* = \omega_0 \Delta T_j/(8\, a \, \sqrt{\alpha} \, k_0^2)$. 

The numerical values of the positions at which the collisions between the
counterpropagating dam break flows occur are summarized in Table I
for the three boxes generated in our experiment. 
The breaking lines separating the genus $0$ region from the genus $1$
region are plotted in Fig. 1(b). It can be readily seen that there is a good
quantitative agreement between theoretical and experimental results.  
In particular, the distance at which the collision is predicted to occur for the largest
box is larger than the physical length of the water tank and it is clear
that the collision between the dam break flows is not experimentally observed 
in this situation, see Fig. 1(b) and also Fig. 4. 

\begin{figure}[h]
  \includegraphics[width=1\linewidth]{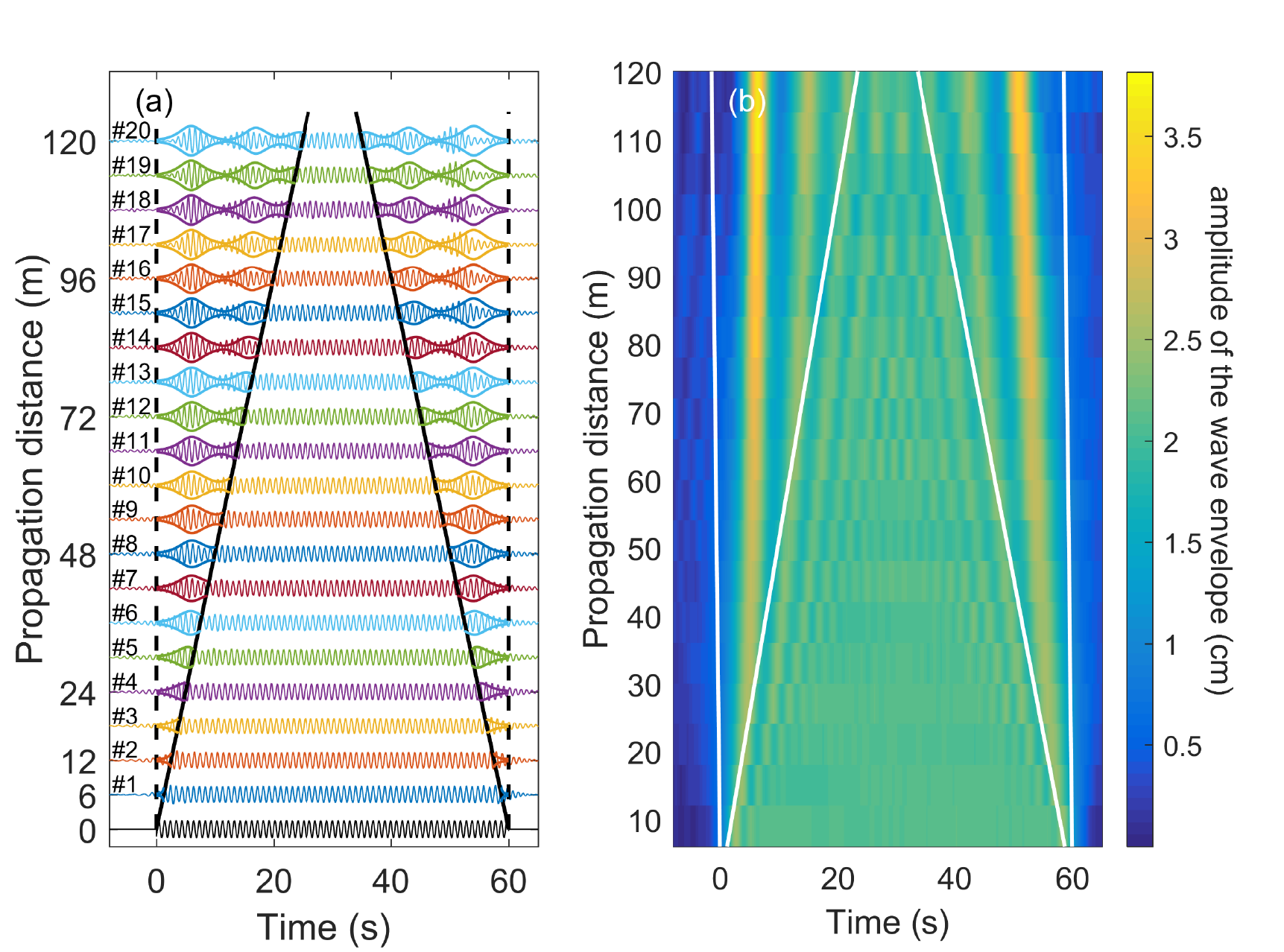}
  \caption{ Experimental results showing the nonlinear evolution
    of the rectangular wave packet of largest size ($\Delta T_3=60$ s, $a=2$ cm,
    $k_0a=0.082$, $T_0=0.99$ s) in Fig. 1(b). (a) Signals recorded by the $20$
    gauges placed all along the tank. All the envelopes superimposed on the
    carrier wave are computed from solutions given by Eqs.
    \eqref{elliptic}, \eqref{mod1} \eqref{mod2} using $\tau_0=-0.48$. The breaking
    lines plotted with full black lines in (a) are computed from Eq. \eqref{Xi1}. (b)
    Space-time evolution of the modulus of the envelope of the experimental signals.
    The white lines in (b)
    represent the breaking curves computed from Eq. \eqref{Xi1}.
    }
\end{figure}

To go one step further in the analysis of our experiment, we now
compare experimental data with the modulated cnoidal solution that has
been discussed in Sec. \ref{SCth}. To this end Eqs. (\ref{elliptic}),
(\ref{mod1}), (\ref{mod2}) are solved and rephrased to physical
variables according to the transformations introduced in Sec. \ref{semi_class}. 
Considering only the box of largest size where the counterpropagating dam
break flows are the most developed near the end of the water tank
($z \sim 120$ m), Fig. 4 shows that the modulated cnoidal wave envelope
determined from the semi-classical theory matches quantitatively well 
the experimental results over the whole 
range of evolution of the dam break flows (i. e. from $z=6$ m
to $z=120$ m). The numerical value of $\tau_0$ has been determined
from the signal measured at the last gauge, at $z_{20}=120$ m ($\tau_0=-0.48$). 
%Note that good quantitative agreement between
%experimental and theoretical results has been obtained for the
%two boxes of smaller sizes, though it is not presented here for the
%sake of simplicity. 

\subsection{Inverse scattering transform analysis of the experimental data}\label{ist}

Some other insights into our experimental results can be obtained
from the perspective of the inverse scattering transform (IST)
method. The configuration
considered in our experiments corresponds to the initial value
problem specified by Eq. (\ref{nlse})  and Eq. (\ref{ic0}).
As shown by Zakharov and Shabat \cite{Zakharov:72},
the nonlinear dynamics in this kind of problem is determined by the IST
spectrum that is composed of two components: a discrete part related to the soliton
content of the box data and of a continuous part related to the dispersive radiation.
In particular, it is known  that the number $N$ of solitons
embedded inside the initial box is given by $N=\hbox{int}(1/2+1/(\pi \epsilon))$,
where $\hbox{int}(x)$ denotes the integer part of $x$
\cite{Manakov:74a,Manakov:74b,Burzlaff:88,Kivshar:89,Sedov:18}.

As shown in Table I, the number of solitons that are embedded inside
the rectangular wave trains is predicted to grow from $N=3$ for the smallest box ($\epsilon=0.095$)
to $N=7$ for the largest box ($\epsilon=0.047$). To check this result
from experimental signals and
to investigate more in depth the integrable nature of the features experimentally
observed, we now consider the non-self-adjoint Zakharov-Shabat eigenvalue problem
\begin{equation}\label{ZS}
\epsilon \frac{d \bf{Y}}{d\tau}=
 \begin{pmatrix}
-i \lambda & \psi_0 \\ -\psi_0^* & i \lambda \\ 
 \end{pmatrix}
 \bf{Y}
\end{equation}
that is associated with Eq. (\ref{nlse}). 
$\bf{Y}(\tau; \lambda,\epsilon)$ is a vector where $\lambda \in \mathbb{C}$ represent the
eigenvalues composing the discrete spectrum associated with the soliton content of
the complex envelope $\psi_0$ measured at some given propagation distance. 
Note that the linear spectral problem (\ref{ZS}) can be identified 
as one half of the Lax pair for Eq. (\ref{nlse}) \cite{yang2010nonlinear}. 

Fig. 5 shows the complex eigenvalues $\lambda$ that are computed from the numerical resolution
of Eq. (\ref{ZS}) made by using the Fourier collocation method described and used e.g. in
ref. \cite{yang2010nonlinear,Randoux:16a,Randoux:18}.
For the sake of clarity only the upper part of the complex plane is
represented but complex conjugate eigenvalues are obviously
obtained from the numerical resolution of Eq. (\ref{ZS}). 
Fig. 5 (left column) shows the complex eigenvalues
computed for the three experimental envelopes (red lines in Fig. 3) measured
at $z_1=6$ m, close to the wavemaker. Remarkably, nearly all of the non-zero eigenvalues
numerically computed are distributed close to the vertical imaginary axis, demonstrating that the
solitons embedded inside the three rectangular wave trains have negligible velocity
at the initial time. In fact, the rigorous semi-classical IST analysis of the box
problem \cite{Jenkins:14} shows that the discrete spectrum is located on the imaginary
axis as $\epsilon \to 0$. 
The number of discrete eigenvalues found from numerical IST
analysis and reported in Fig. 5(a),(c),(e) is in good agreement with results reported in Table I.

\begin{figure}[h]
  \includegraphics[width=0.7\linewidth]{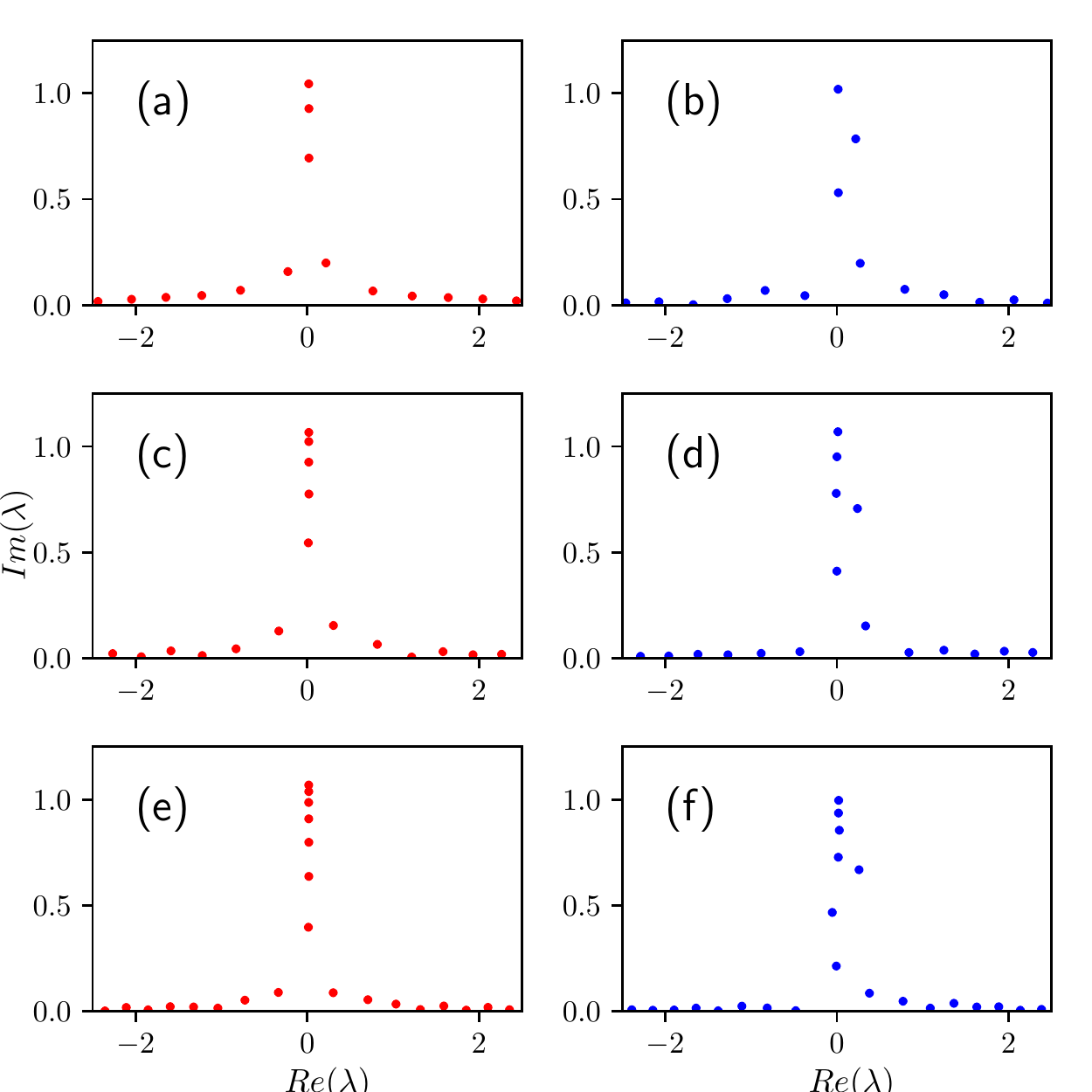}
  \caption{Discrete IST spectra of the three envelopes measured at
    $z_1=6$ m (Left column) and $z_{20}=120$ m (Right column). (a), (b)
    $\Delta T_1 = 30$ s, (c), (d), $\Delta T_2 = 45$ s, (e), (f) $\Delta T_3 = 60$ s.
    Only the upper half part of the complex plane is represented. 
}\label{fig5}
\end{figure}

Fig. 5 (right column) shows the complex eigenvalues
that are computed for the three experimental envelopes (blue lines in Fig. 3) measured
at $z_2=120$ m, far from the wavemaker. In the IST theory of the 1D-NLSE, these
discrete eigenvalues do not change in the evolution time. In the experiment,
we find that this isospectrality condition is not perfectly verified
because of the unavoidable occurrence of small perturbative effects. It is however clear
that the number of eigenvalues is preserved over the propagation distance
characterizing our experiment, i. e. between $z_1=6$ m and $z_2=120$ m. Moreover
the global shape of the IST spectra is well preserved (compare
left and right columns in Fig. 5), thus confirming the nearly integrable nature
of the features observed in the experiment. 

Note that the degree of preservation of the eigenvalues reported in Fig. 5 was
not reached in our initial preliminary experiments because of the occurrence of a
slightly multi-modal propagation in the water tank. Such a multimodal propagation is
prone to occur because the width of the tank ($5$ m) is relatively large as compared
to the typical experimental wavelength ($\sim 1$ m). Therefore we have taken great care
that the motion of the wavemaker effectively produces nearly a single-mode excitation
leading to a dominant 1D wave propagation.

\section{Water wave experiment : Robustness of the observed dynamics to higher-order effects}\label{higher_order_effects}

\begin{figure}[h]
  \includegraphics[width=0.7\linewidth]{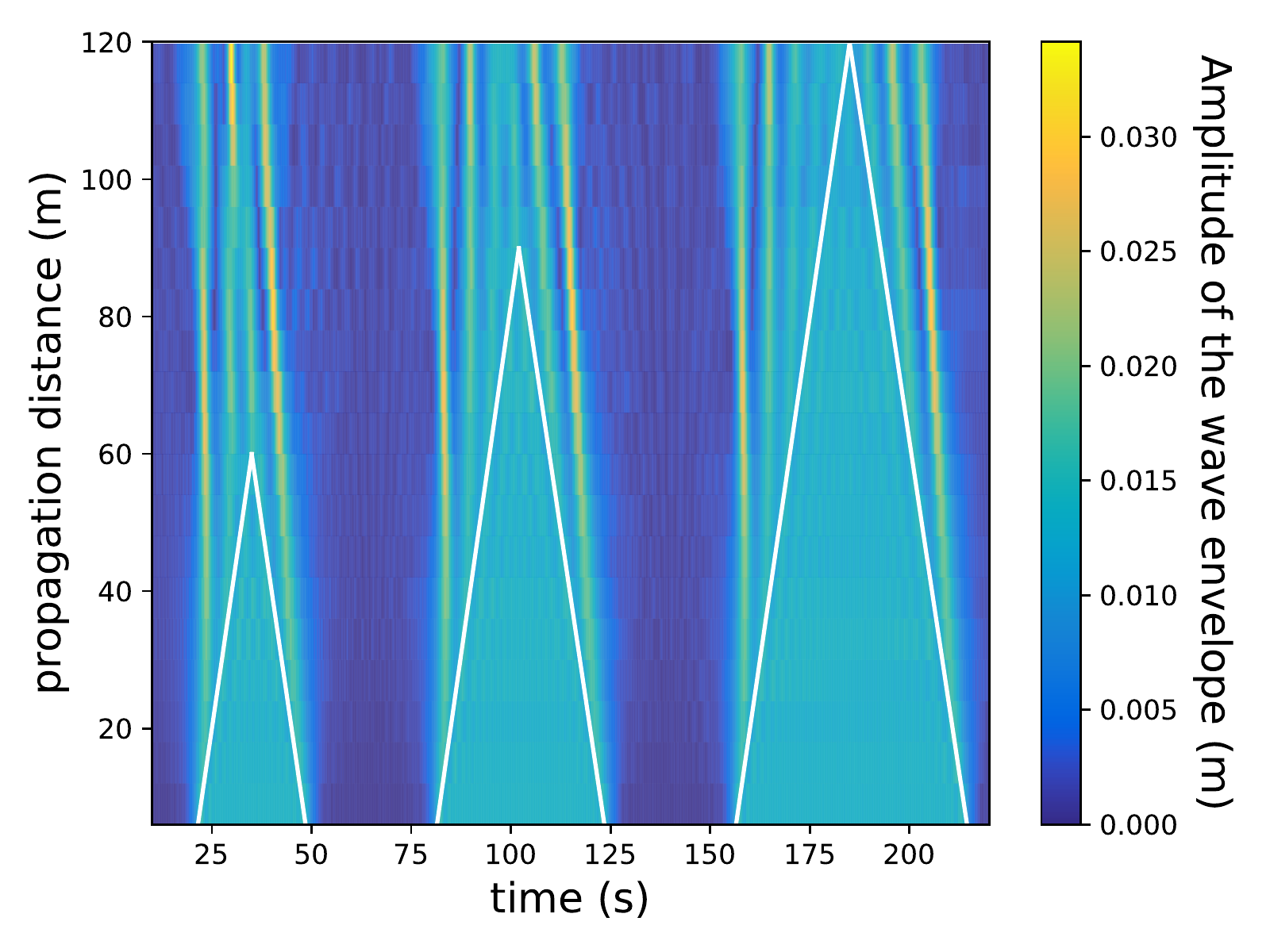}
  \caption{ Space-time evolution of the envelopes of three rectangular wave
    packets ($\Delta T_1=30$ s, $\Delta T_2=45$ s, $\Delta T_3=60$ s) in
    the regime where higher-order nonlinear effects have a perturbative influence.
    The carrier frequency $f_0=1/T_0$ is $1.28$ Hz. Other experimental parameters
    are $k_0=6.58$ m$^{-1}$, $a=0.014$ m (the wave steepness is $k_0 a=0.09$).
    White lines represent breaking curves that are determined using the methodology
    described in Sec. \ref{cnoidal}. 
    }
\end{figure}

\subsection{Space-time evolution}

In this section, we demonstrate that the observed dynamics exhibits some degree
of robustness to higher-order effects that unavoidably perturb the wave evolution
when experimental parameters are changed in such a way that the strength of nonlinearity
increases. To do so, we have simply increased the frequency of the wavemaker from
$f_0=1/T_0=1.01$ Hz to $f_0=1.28$ Hz while also decreasing the
amplitude of the wave envelope from $2$ cm to $1.4$ cm. With these changes, the nonlinear
length decreases from $L_{NL}=39.86$ m to $L_{NL}=18.77$ m while the linear dispersive lengths $L_D$
remain unchanged and identical to those summarized in Table 1.

Fig. 6 shows the space-time evolutions of the three rectangular envelopes that
are observed in this situation where the nonlinearity strength is increased,
see also the Supplemental Material Video S3 \cite{note_supl}. Contrary to the experimental space-time
evolutions considered in Sec. \ref{water_exp} and in Sec. \ref{data_analysis},
there is now a marked asymmetry in the space evolution of the three wave 
packets. Even though frequency down-shifting
is not observed in the studied regime, the asymmetric space-time evolution in Fig. 6
is associated with a significant spectral broadening
phenomenon, see Fourier spectra plotted in the Supplemental
Material \cite{note_supl}. As discussed in details in Sec. \ref{nl_analysis}, integrability
of the wave system is not preserved in this regime where higher-order nonlinear
effects influence the wave dynamics.  In these conditions, the observed dynamics is
not described by the 1D-NLSE but rather by other models like the
unidirectional Zakharov equation or the Dysthe equation \cite{Shemer:02}. 
Note that small higher-order effects are already noticeable in the
details of Fig. 4(a) where the envelope of the modulated cnoidal wave
fits better the left part than the right part of the wave packet
at large distances from the wavemaker.

Despite the undisputable presence of higher-order nonlinear effects in water wave experiments
reported in Fig. 6, it is clear that the scenario of emergence of 
counter-propagating dispersive dam break flows remains qualitatively
well observed. White lines plotted in Fig. 6 represent the breaking
lines that are computed from the semi-classical theory presented in
Sec. \ref{SCth} (see Eq. \ref{Xi1}). At a qualitative level, the breaking lines still clearly separate
regions where DSWs (genus $1$) are found from regions where the (unmodulated)
plane waves (genus $0$) are found. Therefore these breaking lines retain some 
relevance to the description of the dynamics, even in the presence of perturbative
higher-order effects.

\subsection{Nonlinear spectral analysis }\label{nl_analysis}

In the regime where higher-order nonlinear effects influence the dynamics, 
the wave system is no longer described by the 1D-NLSE and rigorously
speaking, the dynamics is no longer of an integrable nature.
However mathematical tools of nonlinear spectral analysis can still
be used to advantage for getting relevant information about the wave
system. For instance, it has been shown in ref. \cite{Randoux:18}
that dissipative effects occurring in a water tank produce some slow modulation
of the spectral (IST) portrait of the Peregrine soliton recorded in water
wave experiments reported in ref. \cite{Chabchoub:11}. More recently the IST has been applied
to characterize coherent structures in dissipative nonlinear systems described
by the cubic Ginzburg-Landau equation \cite{Chekhovskoy:19}. 

Here, we apply nonlinear spectral analysis to examine
the soliton content of the rectangular wave packets in the propagation regime
displayed in Fig. 6. For the sake of simplicity, we only present here
the numerical results that are associated with the box of duration $\Delta T_2=45$ s
(central box in Fig. 6).
As described in Sec. \ref{ist} and also more in details in
ref. \cite{Randoux:16a,Randoux:18}, the determination of the discrete IST eigenvalues
relies on the numerical resolution of Eq. (\ref{ZS}) for the potentials $\psi_0$
that are measured in the experiment. 

Fig. 7(a) shows the rectangular envelope of the central box of Fig. 6 that has been measured
at $z_1=6$ m, close to the wavemaker. Fig. 7(b) shows the corresponding
discrete IST spectrum which is composed of $7$ eigenvalues located
well above the real axis. Let us recall that the discrete IST spectrum
of the same box was only composed of $5$ eigenvalues in the regime
where the dynamics was described by the integrable focusing 1D-NLSE,
see Sec. \ref{ist}. The result of an increased nonlinearity 
is therefore that the number of solitons embedded within the box has
increased, which is not that surprizing but which is here substantiated
and quantified with the IST. 

\begin{figure}[h]
  \includegraphics[width=0.7\linewidth]{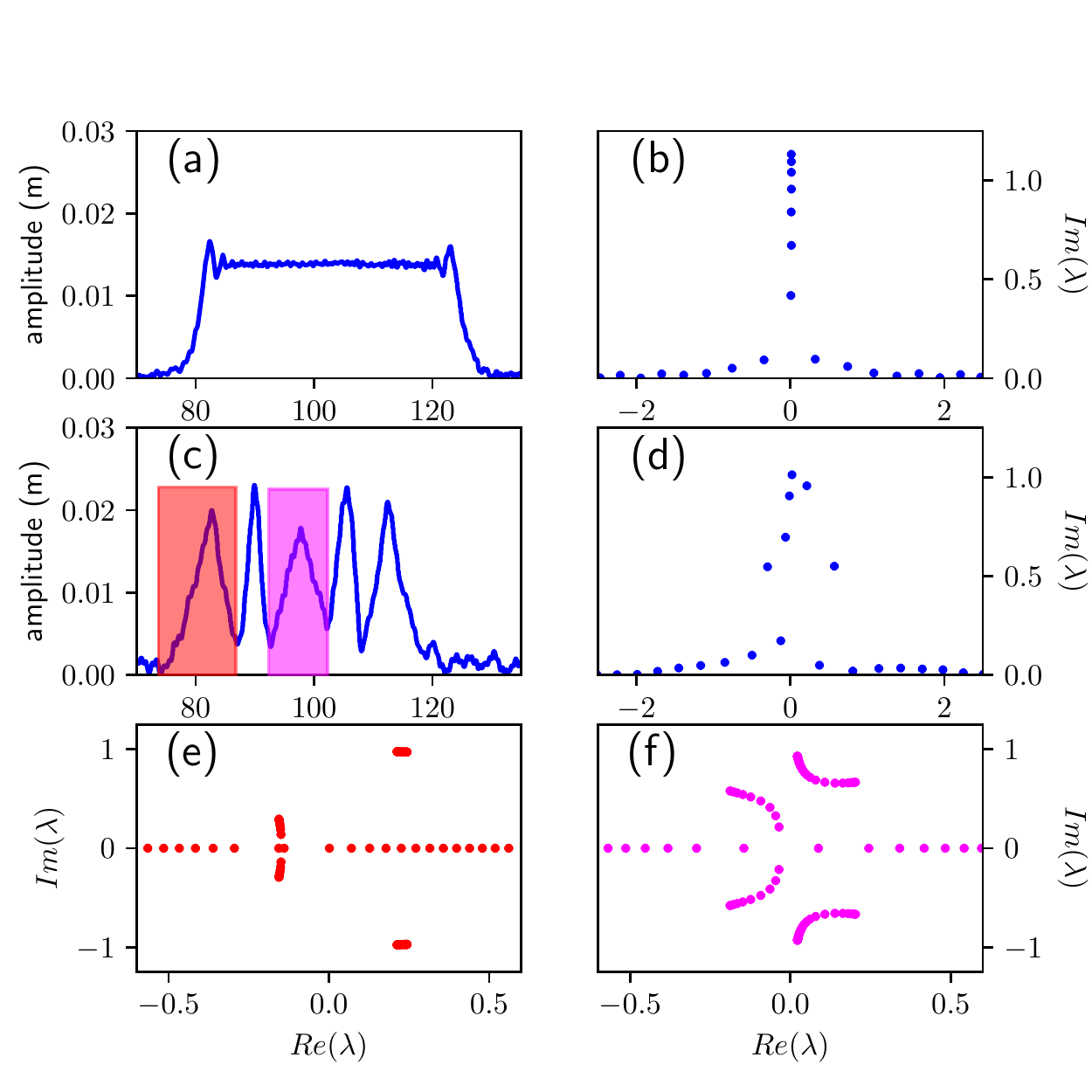}
  \caption{(a) Envelope of the central wavepacket of Fig. 6 measured at $z_1=6$ m,
    close to the wavemaker. (b) Discrete IST spectrum of the wavefield
    plotted in (a). (c) Envelope of the central wavepacket of Fig. 6 measured
    at $z_{20}=120$ m, far from the wavemaker. (d) Discrete IST spectrum of the wavefield
    plotted in (c). (e) Local IST spectrum of the coherent structure highlighted in
    red in (c). (f) Local IST spectrum of the coherent structure highlighted in
    magenta in (c).
    }
\end{figure}

Fig. 7(c) shows the envelope of the central box of Fig. 6 that has been measured
at $z_{20}=120$ m, far from the wavemaker. Fig. 7(d) shows the corresponding
discrete IST spectrum. Comparing Fig. 7(d) and Fig. 5(d), we obtain the clear
signature that higher-order effects significantly perturbate the discrete IST
spectrum (i. e. the soliton content of the rectangular wave packet). 
Six eigenvalues well above the real axis are observed instead of
seven near the wavemaker, see Fig. 7(a). Moreover the
real parts of most of these eigenvalues become nonzero which means that
the solitons embedded within the box have acquired some velocity, a feature
that is fully compatible with the fact that the rectangular box exhibits some slow drift
in the space-time plot, see Fig. 6. 

To investigate the change of the genus \cite{GEl:16} of the coherent 
structures emerging in the space-time evolution shown in Fig. 6, we have used the tools of local IST analysis
introduced in ref. \cite{Randoux:16a} and already applied for the
analysis of experimental signals in ref. \cite{Randoux:18}. 
In the approach used for local IST analysis, the analyzed coherent structure
is isolated by truncating the wave profile over some given time interval.
The truncated wave field is then periodized in time. This produces some
local finite-band approximation of the wave field which can be interpreted within
the framework of finite gap theory \cite{Tracy:84,Grinevich:18,osborne2010nonlinear}.
Numerically solving
Eq. (\ref{ZS}) for the periodized potential $\psi_0$, we obtain a spectrum
that is composed of bands. The number $N$ of bands
composing the nonlinear spectrum determines the genus $g=N-1$ of the solution
that can be viewed as a measure of complexity of the space-time evolution of
the considered solution. 

Fig. 7(e) shows the local IST spectrum of the peaked structure highlighted in red in
Fig. 7(c) and located at the left-edge of the box.
Rigorously speaking the spectrum is composed of $3$ bands but
the small-amplitude band crossing the real axis can be seen as being pertubative. Therefore
the local IST spectrum can be seen as being mainly composed of two complex conjugate bands,
confirming the nearly genus $1$ nature (soliton like) of the structure analyzed
at the left edge of the box. Note that the spectrum plotted in Fig.~7(e) is
qualitatively very similar to the spectrum computed in Fig.~2(e) from
numerical simulations of the focusing 1D-NLSE. 

Fig.~7(f) shows the local IST spectrum of the peaked structure highlighted in magenta in
Fig.~7(c) and located in the center of the box. This spectrum is composed
of $3$ main bands confirming that the observed object represents a coherent structure
of genus $2$ (i.e. of breather type). Note that the spectrum plotted in Fig.~7(f) is
qualitatively similar to the spectrum computed in Fig.~2(f) from
numerical simulations of the focusing 1D-NLSE. Contrary to Fig. 2(f),
the spectrum of Fig.~7(f) presents a marked asymmetry that we interpret as
arising from the higher-order effects that perturbate the integrable dynamics. 
\\

\section{Summary and Conclusion}\label{conclusion}

In this paper, we have reported experiments showing the evolution of
nonlinear deep-water surface gravity waves having their initial envelopes in the form
of large-scale near-rectangular barriers. We have shown that nonlinear wave packets
are not necessarily disintegrated by the Benjamin-Feir instability and that
there exist some regimes in which a specific, strongly nonlinear modulation,
propagates from the edges of the wavepacket towards the center with finite
speed. The observed counter-propagating dispersive dam break flows  represent modulated nonlinear wave trains  that can be described within the framework of the semi-classical 1D-NLSE. They could be viewed as examples of DSW dynamics persisting in focusing (modulationally unstable) nonlinear media.

Our experimental results are shown to be in good quantitative agreement 
with predictions of the 1D-NLSE semi-classical theory \cite{Jenkins:14, GEl:16}, confirming
the robustness of the observed dynamical scenario  with respect to
perturbative higher-order nonlinear effects inevitably present in a water wave experiment. 
We have also shown that nonlinear spectral analysis \cite{Randoux:18} can be used to advantage
to determine the soliton content of the generated wavepackets while also
providing useful information about the  local wave dynamics in terms of
the number of fundamental nonlinear wave modes (the genus) comprising the
observed structure at a given space-time point. 

By confirming that DSW dynamics can be observed in deep water waves, 
our work opens way to further experimental and theoretical investigations on
the subject of DSW formation in focusing nonlinear media. In particular
several scenarios \cite{Biondini:18} associated with the so-called Riemann problems---i. e. 
the evolution of a jump discontinuities (not necessarily dam breaks) between two uniform states of the
initial field---could be also possibly observed in deep water waves. 

Interesting questions are related to the competition between the
DSW formation and the Benjamin-Feir instability. Our experiments are performed in a
regime where the DSW dynamics plays the dominating role and the effects of the Benjamin-Feir instability can be neglected, but it would be  interesting  to examine in detail how the Benjamin-Feir
instability affects the DSW structure at longer propagation times. Finally, our experiments have shown
that the generated DSWs exhibit certain robustness to higher-order nonlinear effects.
It is another interesting and challenging question to investigate these higher order effects
more in detail from the theoretical perspective. 

\appendix

\section{Finite depth effects}\label{appendixa} 

As shown in ref. \cite{Mei:92}  the weakly nonlinear, narrow-banded
approximation of the fully nonlinear irrotational and inviscid water wave equations
is the 1D-NLSE under the following form 
\begin{equation}\label{nlse_phys_full}
\frac{\partial A}{\partial t} + \frac{1}{2} \frac{\omega_0}{k_0} \nu \frac{\partial A}{\partial z} +
i \frac{1}{8} \frac{\omega_0}{k_0^2} \kappa \frac{\partial^2 A}{\partial z^2}
+ i \frac{1}{2} \omega_0 \, k_0^2 \, \gamma |A|^2 A=0 ,  
\end{equation}
where $A$ is the complex envelope of the water wave. $\nu$ is the correction to the
group velocity for finite depth. $\kappa$ and $\gamma$ are coefficients that
in general depend on the water depth $h$ at the dominant wave number $k_0$
and at the corresponding angular frequency $\omega_0$.

The general expressions of $\nu$, $\kappa$ and $\gamma$ are given by
(see e.g. ref. \cite{Shemer:15} and see ref. \cite{Mei:92} for the derivation)
\begin{equation}\label{nu}
\nu=1+ \frac{2k_0 h}{\sinh{(2k_0 h)}}
\end{equation}
\begin{equation}\label{kappa}
\kappa=-\nu^2+2+8(k_0h)^2 \frac{\cosh{(2k_0 h)}}{\sinh^2{(2k_0 h)}}
\end{equation}
\begin{equation}\label{gamma}
  \begin{split}
  \gamma= \frac{\cosh{(4k_0 h)}+8-2\tanh^2{(k_0h)}}{8\sinh^4{(k_0 h)}} \\
  - \frac{(2\cosh^2{(k_0 h)}+0.5\nu)^2}{\sinh^2{(2k_0 h)}} \left(\frac{k_0h}{\tanh{(k_0h)}}-\frac{\nu^2}{4} \right)
  \end{split}
\end{equation}

For a hydrodynamic wavemaker problem, it is convenient to use the 1D-NLSE
in the form of an evolution equation in space. Using changes of variables
described in ref. \cite{osborne2010nonlinear,Chabchoub:16},
one obtains the following evolution equation
\begin{equation}\label{nlse_phys_correc_finite}
i \frac{\partial A}{\partial z} + \frac{\kappa}{\nu^3} \frac{k_0}{\omega_0^2}
\frac{\partial^2 A}{\partial t^2} + \frac{\gamma}{\nu} k_0^3 |A|^2 A=0 ,  
\end{equation}
in the frame of reference moving with the group velocity of the wave packets.

In the experiments reported in Fig. 1(b), the numerical value of $k_0 h$ is $12.3$ and
the numerical values of the corrective terms $\nu$ and $\kappa$ given by
Eqs. (\ref{nu}) and (\ref{kappa}) are
very close to unity. However the numerical value of $\gamma$ is $\sim 0.91$ which
means that the finite-depth correction to the cubic nonlinearity is small but
not negligible. Therefore our experiments are described by
Eq. (\ref{nlse_phys_correc_finite}) in which the values of the corrective
terms are set to $\kappa=\nu=1$ and $\gamma=\alpha=0.91$.

\begin{acknowledgments}
This work has been partially supported  by the Agence Nationale de la
Recherche  through the LABEX CEMPI project (ANR-11-LABX-0007),   the
Ministry of Higher Education and Research, Hauts de France council and
European Regional Development  Fund (ERDF) through the Nord-Pas de
Calais Regional Research Council and the European Regional Development
Fund (ERDF) through the Contrat de Projets Etat-R\'egion (CPER
Photonics for Society P4S). The work of GE was partially supported by
EPSRC grant EP/R00515X/2. The work of FB, GD, GP,
GM, AC and EF was supported by the French National Research Agency
(ANR DYSTURB Project No. ANR-17-CE30-0004). EF thanks partial support
from the Simons Foundation/MPS N$^{\rm o}$651463.
\end{acknowledgments}

%\bibliography{DB}

%merlin.mbs apsrev4-1.bst 2010-07-25 4.21a (PWD, AO, DPC) hacked
%Control: key (0)
%Control: author (0) dotless jnrlst
%Control: editor formatted (1) identically to author
%Control: production of article title (0) allowed
%Control: page (1) range
%Control: year (0) verbatim
%Control: production of eprint (0) enabled
%

\end{document}